\documentclass[journal,twocolumn,twoside]{IEEEtran}

\usepackage{epsfig,amsmath,amssymb,epsf,amsthm,scalefnt,multirow,subfig}
\usepackage{caption}
\usepackage{xcolor}
\usepackage{float}
\usepackage{cite}
\usepackage{psfrag}
\usepackage{cleveref}
\usepackage{mathtools}
\usepackage{lipsum}
\usepackage{xcolor}

\newtheorem{lemma}{Lemma}
\newtheorem*{lemma*}{Lemma}

\theoremstyle{definition}

  \def\cC{{\mathcal{C}}} 
   
   \def\cL{{\mathcal{L}}}
 \def\cN{{\mathcal{N}}}

\def\Re{\mathop{\mathrm{Re}}}
\def\Im{\mathop{\mathrm{Im}}}

\def\b0{{\pmb{0}}} 

\def\bE{{\mathbf{E}}}

\begin{document}

\title{Practical Distributed Reception for Wireless Body Area Networks Using Supervised Learning
\thanks{J. Cha and J. Choi are with the School of Electrical Engineering, Korea Advanced Institute of Science and Technology (e-mail: {charge, junil}@kaist.ac.kr).}
\thanks{D. J. Love is with the School of Electrical and Computer Engineering, Purdue University (e-mail: djlove@purdue.edu).}}

\author{Jihoon~Cha, Junil~Choi, and David~J.~Love}

\maketitle

\begin{abstract}
Medical applications have driven many areas of engineering to optimize diagnostic capabilities and convenience. In the near future, wireless body area networks (WBANs) are expected to have widespread impact in medicine. To achieve this impact, however, significant advances in research are needed to cope with the changes of the human body's state, which make coherent communications difficult or even impossible. In this paper, we consider a realistic noncoherent WBAN system model where transmissions and receptions are conducted without any channel state information due to the fast-varying channels of the human body. Using distributed reception, we propose several symbol detection approaches where on-off keying (OOK) modulation is exploited, among which a supervised-learning-based approach is developed to overcome the noncoherent system issue. Through simulation results, we compare and verify the performance of the proposed techniques for noncoherent WBANs with OOK transmissions. We show that the well-defined detection techniques with a supervised-learning-based approach enable robust communications for noncoherent WBAN systems.
\end{abstract}


\section{Introduction}\label{sec1}
\IEEEPARstart{I}NCREASES in life expectancy and investments in healthcare technology have led to advances in sophistication and integration in medical-use electronic devices. This has been particularly evident in sensing devices that hinge on wireless connectivity. Wireless body area networks (WBANs), which are standardized through the IEEE 802.15.6 task group, hold much promise in increasing the prevalence and acceptance of wearable devices \cite{IEEE802.15.6,Movassaghi:2014,Latre:2011,Shinagawa:2004}. At the same time, WBANs could potentially be used in non-medical applications such as gaming devices or mobile applications that respond to physical activities \cite{Fortino:2012}. For WBANs to become widely used, however, multiple challenges in power consumption, form factor, and user-oriented design must be overcome \cite{Cotton:2014,Hanson:2009}. Recent works take these challenges into account for various applications, e.g., group activity management, remote patient monitoring, and interference management \cite{Liu:2020,Niaz:2020,Mehmood:2020}. Comprehensive research, though, is still required to implement practical wireless devices optimized for the ever-increasing understanding of the human body's electromagnetic (EM) characteristics.

Three physical layers (PHYs) for WBANs are defined by the IEEE 802.15.6 task group: narrowband (NB), ultra-wideband (UWB), and body channel communication (BCC) \cite{Cavallari:2014}. When wireless communication is conducted inside or on the human body using radio frequency (RF) bands, the body itself significantly influences the communication channel. Unfortunately, WBAN propagation through the body does not follow the traditional over-the-air channel models. NB and UWB signals propagate both through the surface of the skin and the air, and they can suffer from a large path loss due to body blockage \cite{Fort:2006}. This interaction with the human body is regulated and leads to heating of the tissue \cite{Hochwald:2014,Ying:2015,Ebadi-Shahrivar:2019,Castellanos:2020}.
In NB and UWB PHYs, several efforts to statistically characterize the channel in such a non-traditional environment were conducted in \cite{Sangodoyin:2018,Ambroziak:2016,Smith:2011}, and the theoretical performance, including power allocation, channel capacity, and outage probability, was investigated in \cite{Razavi:2019,Cheffena:2015}.

The BCC systems operate in the carrier frequencies of 5-50 MHz \cite{Cotton:2014}, utilizing EM propagation through both the skin surface and the entire human body for communication. The electrical conductivity of human tissue is higher than that for air, and a lower power consumption of transceivers for the BCC systems is guaranteed by using an electrode \cite{Bae:2012}. The properties of the human body imply that the channel experienced by BCC can vary greatly across the different frequencies. The channel for the BCC PHY was modeled as a simple electronic circuit, which operates based on capacitive and galvanic coupling in \cite{Seyedi:2013,Callejon:2012,Bae:2012}. In terms of wireless communications, there has been relatively little work on stochastic channel modeling for BCC since a larger number of parameters that influence the channel conditions make it difficult to classify and analyze the channel.

The difficulty in modeling the WBAN channel is also compounded by its quickly varying characteristics. Properties of the human body along with its movement can lead to a channel that is not easily tracked. For this reason, WBANs will often have to operate noncoherently, i.e., all transceivers can acquire neither instantaneous nor statistical channel state information (CSI) due to the 1) fast varying body state such as temperature, posture, and composition in time, and 2) varying channel statistics depending on people \cite{Sangodoyin:2018,Ambroziak:2016,Smith:2011}. In such a situation, it would be preferable to use only the envelope of the instantaneous received signals for symbol detection. There are many previous works, including \cite{Jing:2016,Hammouda:2015}, on communication systems using the envelope of the received signals; however, most of them exploited channel statistics at the receiver side, which is not suitable for WBANs.

In wireless communications, receivers can detect data symbols by defining decision regions. This is fundamentally a problem of classification, and the classifier can usually be designed using pilot symbols. As noted above, however, it is difficult to accurately characterize the input-output model used for WBAN communication and thereby to conduct classification as in a conventional over-the-air communication. This difficulty of accurate channel modeling motivates the use of machine-learning techniques that can leverage empirical data with training samples derived from an unknown system. In this paper, we focus on supervised learning to detect data symbols. Supervised-learning approaches have been used for data detection \cite{Jeon:2018} and adaptive modulation \cite{Daniels:2010} in wireless communications. However, these past techniques are only valid for channels with long coherence times.

We also explore how diversity techniques can be used to overcome the challenging propagation conditions in WBANs. Distributed reception employs multiple receiving nodes that are geographically separated \cite{Choi:2015,Choi:2015b,Ibrahim:2017,Brown:2014}. These nodes usually guarantee a low cost and low power consumption, and moreover, the rudimentary data acquired from each node can be jointly used to facilitate performance approaching that of a centralized system. A wearable communication system with distributed reception over-the-air was also examined in \cite{Ouyang:2009}, but distributed reception has not been applied on noncoherent WBAN systems.

One issue with WBANs is that the channel characteristics can behave unpredictably due to the fluid, movement, and heterogeneous composition of the human body. In this paper, we address this by proposing a noncoherent WBAN transmission and reception framework. Distributed reception, constrained using practical assumptions, is exploited in this system to obtain spatial diversity. To the best of our knowledge, this paper proposes for the first time ever noncoherent distributed detection, while most (if not all) previous works on wireless distributed detection are based on at least some form of coherent processing \cite{Choi:2015,Choi:2015b,Ibrahim:2017,Brown:2014,Ouyang:2009,Park:2014,Yang:2011}. We focus on the popular and standardized technique of on-off keying (OOK) for signaling, which is a binary modulation and one of the mandatory modes in certain frequency bands for WBANs \cite{IEEE802.15.6,Zhao:2017}. We adopt distributed reception to reliably detect OOK symbols where a fusion center, which is wired with geographically separated receiving nodes, collects necessary information for the symbol detection. A preliminary study was performed in \cite{Cha:2020} with an idealistic assumption of real-valued~channels.

We propose two novel symbol detection approaches for noncoherent WBANs where both approaches are not able to utilize conventional channel estimation. One approach is based on marginalization in probability where the fusion center computes marginal distributions from the joint probability distribution of the received signals in an empirical manner. This probabilistic noncoherent detection approach utilizes the entire received pilot signals for each time slot to compute these probability distributions for the symbol detection, resulting in huge computational overhead.

The second proposed symbol detection approach is based on supervised learning. This approach is developed by relying only on the traditional pilot and data transmission framework used in most wireless communication systems. Pilot symbols are used to compute a small amount of refined information called reference values for data symbol detection. In contrast to conventional symbol detection, our supervised-learning-based approach copes with a totally unknown WBAN system at transceivers where the channels are time-varying, and static channel knowledge acquisition is of little use for detection. We develop three symbol detection techniques for the supervised-learning approach without adopting any complex deep learning algorithm. Employing only a few reference values, these techniques make the symbol detection process much simpler with less complexity than that of the marginalization-based approach.

We verify through simulations with realistic body channel models that the supervised-learning-based approach is comparable or even superior to the marginalization-based approach. The supervised-learning-based approach can even make comparable performance to the case using perfect CSI for coherent detection.

The paper is organized as follows. Our system model employing OOK modulation for WBANs is described in Section \ref{sec2}. We explain the first OOK symbol detection approach based on marginalization, called empirical likelihood ratio test (eLRT), in Section \ref{sec3}. In Section \ref{sec4}, the second detection approach based on supervised learning is introduced where three detection techniques, namely, weight-comparing noncoherent detection (WCNDe) techniques with \mbox{probability-,} deviation-, and combination-valued weights, are proposed in detail. The practicality of WCNDe is also discussed in the section. Simulation results for these detection approaches are shown in Section \ref{simul}, followed by the conclusion in Section~\ref{conc}.

\textbf{Notations:} $ f_{a[n]}(a) $ represents the probability density function for the random process $ a[n] $ where $ n $ is the time instant. $ \bE_{a[n]}\{b[n]\} $ denotes the expectation over $ a[n] $ of the random process $ b[n]$. $ \lvert \cdot \rvert $ denotes the absolute value of a complex number. The function $ \cC\cN\left(\mu,\sigma^2\right) $ is the complex-valued normal distribution function with the mean $ \mu $ and the variance $ \sigma^2 $. $ \Re(z) $ and $ \Im(z) $ represents the real and imaginary parts of a complex value $ z \in \mathbb{C}$ where $ \mathbb{C} $ is the set of all complex~numbers.

\section{System Model}\label{sec2}
\begin{figure}
	\centering
	\includegraphics[width=0.7\columnwidth]{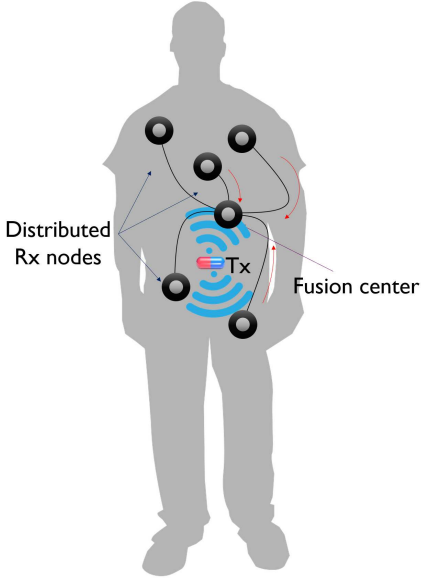}
	\caption{Structure of a WBAN SIMO system. Distributed receiving nodes convey the information of received signals to the fusion center to detect OOK symbols.} \label{figure1}
\end{figure}

We model a single-input multiple-output (SIMO) WBAN system with a transmitter and $ K $ geographically separated receiving nodes as in Fig. \ref{figure1}. The fusion center is wired with the receiving nodes to collect necessary information for symbol detection. The channel between the transmitter and each of the receiving nodes depends on where the transceivers are located and through which path transmitted signals are conveyed in a human body. Therefore, the channel between the transmitter and each receiving node follows a unique probability distribution. The channel distribution is influenced by the complicated structure of the human body, including blood fluid or posture, and its channel gain continuously varies with time due to the rapid change of the body's state. Consequently, these factors make it impossible for the transceivers to obtain instantaneous or statistical CSI.

For time slot $ n $, the received signal at the $ k $-th receiving node is given as
\begin{align}\label{input_output}
  y_k[n] = \sqrt{P}h_k[n] x[n]+w_k[n],
\end{align}
where $ x[n]\in\{0,1\} $ is the transmit symbol that represents OOK modulation, the simplest form of amplitude shift keying and $ P $ is the transmit power. The complex-valued channel gain of the $ k $-th receiving node $ h_k[n] \in \mathbb{C}$ follows a certain circularly-symmetric probability distribution $ f_{h_k[n]}(h_k) $, i.e., $ h_k[n] \sim f_{h_k[n]}(h_k) $. The channel distribution for each $ k $ is assumed to be stationary, i.e., $ f_{h_k[n]}(h_k)=f_{h_k[m]}(h_k) $ for all time instances $ n $ and $ m $. In other words, the instantaneous channel gain varies in each time slot, but its channel distribution does not change during the time of interest. The noise $ w_k[n]\in \mathbb{C}$ is circularly-symmetric, independent, and identically distributed following $ \cC\cN\left(0,N_0B\right) $ where $ N_0 $ and $ B $ are respectively the noise spectral density and the bandwidth. We assume all $ K $ channels and noises are independent of each~other.

We assume that CSI is unknown to the transceivers, which means that each receiving node has no knowledge of $ h_k[n] $ and $ f_{h_k[n]}(h_k) $ at all. Although there are some theoretical channel models for WBANs, e.g., \cite{Sangodoyin:2018,Ambroziak:2016,Smith:2011,Razavi:2019}, they differ depending on a person and even the location within a person. Therefore, it is desirable to develop detectors that can work for arbitrary channel models. In this paper, we implement symbol detectors without using any knowledge of instantaneous or statistical CSI while we adopt body channel models from \cite{Razavi:2019} for numerical simulations in Section \ref{simul}.

Even though the channel characteristics of the WBAN system make acquisition of instantaneous CSI infeasible, there is still benefit to using pilots. We assume that the transmitter sends a known pilot sequence whose length is $ N_p $ as
\begin{align}\label{training_signal}
x[n] = \begin{dcases}
1, &\text{for } n=1,2,\cdots,\frac{N_p}{2}, \\
0, &\text{for }n=\frac{N_p}{2}+1,\cdots,N_p,
\end{dcases}
\end{align}
which is equally applied to all of the proposed approaches in this paper. Sending predetermined OOK pilot symbols of $ 1 $ and $ 0 $ would not give the fusion center instantaneous CSI, but it would provide some empirical channel statistics for noncoherent OOK symbol detection, which will be explained in the next two sections. After $ N_p $ pilot transmissions, we assume the transmitter sends each OOK data symbol equally likely for $ N_d $ channel uses throughout the paper.

\section{Marginalization-Based OOK Symbol Detection Approach}\label{sec3}
In this section, we propose a detection approach that is similar to a detection criterion with statistical CSI in which the pilot symbols are employed for empirical marginalization of a joint probability distribution. This approach is motivated by estimation of a probability density function \cite{Parzen:1962}. In general, statistical knowledge of channel input-output relationships is given to a receiver and utilized to design a detector. Probability distributions for two or more hypotheses from a certain symbol constellation would be the common knowledge used to conduct LRT for symbol detection. However, we assume the transceivers for WBANs do not know any instantaneous CSI due to the fast-varying channels of a human body or any statistical CSI because of the uncertainty of the body's movements and node locations. Instead, we will exploit estimated empirical probability distributions from the received pilot signal sequence in this section.

We start from the ideal case when the true channel distributions are known at the receiver side. With the channel distribution $ f_{h_k[m]}(h_k) $ at each receiving node, symbol detection through LRT is conducted with the conditional distribution of the received signal assuming a specific transmit symbol, computed as 
\begin{multline}\label{conddis}	
f_{y_k[n]\mid x[n]}\bigl(y_k\big\vert x\bigr)=\\\int_{\mathbb{C}} f_{h_k[m]}(h_k)f_{y_k[n]\mid x[n],h_k[m]}\bigl(y_k\big\vert x,h_k\bigr) \mathrm{d}h_k,
\end{multline}
where it is implicitly assumed that the channel distribution is independent of the transmit symbol.
We write the likelihood function for statistical CSI as
\begin{align}\label{marg}
\cL_{\mathrm{stat}}\bigl(x[n]\big\vert y_1[n],\cdots,y_K[n]\bigr)= \prod_{k=1}^K{f_{y_k[n]\mid x[n]}\bigl(y_k\big\vert x\bigr)},
\end{align}
for $ x[n]=1 $ or $ x[n]=0 $. Note that this expression is normalized using the assumption that the two binary OOK symbols are equally likely. The transmit symbol is finally detected by conducting LRT with \eqref{marg} for the two hypotheses. This ideal case, however, is not possible when the receiver does not know the channel distributions.

Alternatively, we introduce the detection technique, called eLRT, to compute an empirical likelihood function using the pilot symbols in \eqref{training_signal}. During the data transmissions, using the pilot signal sequence received during the pilot transmissions and an instantaneous received data signal, the fusion center conducts LRT. The likelihood functions are defined as
\begin{multline}\label{eLRT1}
\cL_{\mathrm{eLRT}}\bigl(x[n]=1\big\vert  y_1[n],\cdots,y_K[n]\bigr)=\\
\prod_{k=1}^K{\frac{2}{N_p}\sum_{m=1}^{{N_p}/{2}}\frac{c}{\pi}e^{-c\lvert y_k[n]-y_k[m]\rvert^2}},
\end{multline}
and
\begin{multline}\label{eLRT0}
\cL_{\mathrm{eLRT}}\bigl(x[n]=0\big\vert  y_1[n],\cdots,y_K[n]\bigr)=\\
\prod_{k=1}^K{\frac{2}{N_p}\sum_{m={N_p}/{2}+1}^{N_p}\frac{c}{\pi}e^{-c\lvert y_k[n]-y_k[m]\rvert^2}},
\end{multline}
for $ N_p+1 \leq n \leq N_p+N_d $ channel uses. The function $ {\frac{c}{\pi}}e^{-c\lvert y_k[n]-y_k[m]\rvert^2} $, which integrates to one over $ \mathbb{C} $, converges to a Dirac delta function as $ c\rightarrow\infty $ and its variance goes to zero. The sample average of the function $ {\frac{c}{\pi}}e^{-c\lvert y_k[n]-y_k[m]\rvert^2} $ could be considered as a smoothed continuous version of a probability mass function that converges to the true probability density function as $ N_p,c\rightarrow\infty $ in probability \cite{Parzen:1962}. This is explicitly derived in the following lemma.
\begin{lemma}\label{conv}
The empirical function $\cL_{\mathrm{eLRT}}(\cdot) $ converges in probability to the likelihood function with only the statistical CSI $ \cL_\mathrm{stat}(\cdot) $, i.e.,
\begin{multline}\label{eLRT_to_marg}
\cL_{\mathrm{eLRT}}\bigl(x[n]\big\vert  y_1[n],\cdots,y_K[n]\bigr)\\ \overset{p}{\longrightarrow}
\cL_{\mathrm{stat}}\bigl(x[n]\big\vert  y_1[n],\cdots,y_K[n]\bigr),
\end{multline}
as $ N_p,c\rightarrow \infty $ for both cases of $ x[n]=1 $ and $ 0 $.
\end{lemma}
Please see Appendix \ref{proof_conv} for the proof.

The likelihood functions of eLRT in \eqref{eLRT1} and \eqref{eLRT0} resemble \eqref{marg}, but do not exploit statistical CSI. Although performance of eLRT with large $ N_p $ and $ c $ approaches to that with the statistical CSI, it requires enormous pilot overhead to converge. Moreover, compared to the techniques to be proposed in the following section, eLRT is inefficient in that the entire received pilot signal sequence $ y_k[m] $ for $ m=1,\dots,N_p $ is used to compute $\cL_{\mathrm{eLRT}}\bigl(x[n]\big\vert  y_1[n],\cdots,y_K[n]\bigr) $ for symbol detection in each time slot.

\section{Supervised-Learning-Based OOK Symbol Detection Approach}\label{sec4}
Using the received pilot signals labeled as 1 or 0, the fusion center can learn weights that work as kernel functions to classify the received data signals. This learning process for symbol detection can be considered as a kind of model-based classical machine learning belonging to supervised learning, which is different from deep learning based on complex learning algorithms and excessive computational resources. Compared to eLRT using the entire pilot symbols, the proposed approach employs only reference values computed during the pilot transmissions, which reduces computational complexity significantly. Three kinds of detection techniques based on supervised learning, i.e., probability-WCNDe (p-WCNDe), deviation-WCNDe (d-WCNDe), and combination-WCNDe (c-WCNDe), are proposed in this section. Among them, c-WCNDe attains the best symbol detection performance by combining the first two WCNDe techniques. We first explain a basic symbol detection rule and then elaborate in detail on the three WCNDe techniques, followed by comments on the practicality of the proposed detection techniques.

\subsection{General detection framework}\label{sec4-1}
\begin{figure}
	\centering
	\subfloat[][]{\includegraphics[width=0.95\columnwidth]{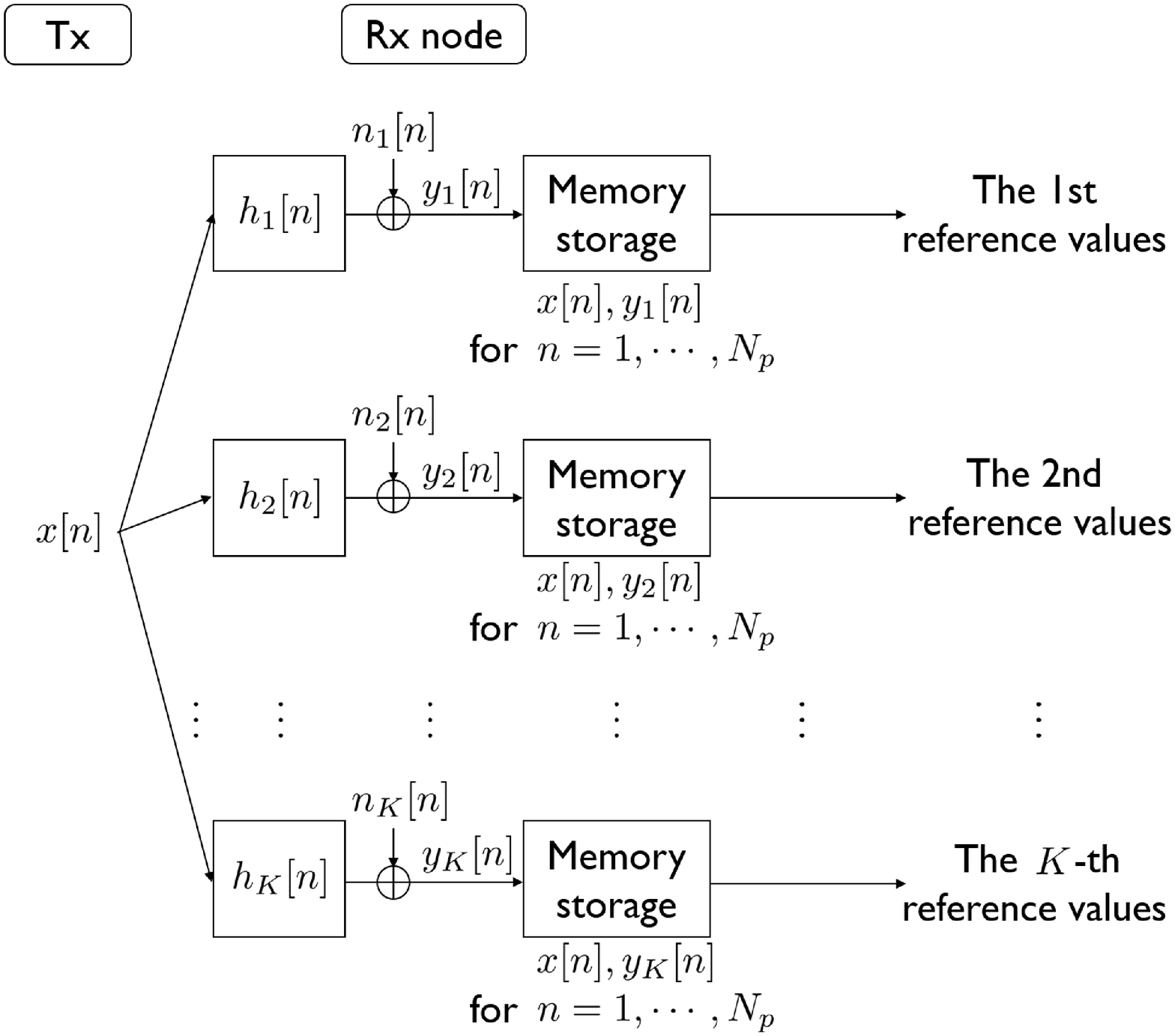}
		\label{training}}
	\hfil
	\subfloat[][]{\includegraphics[width=0.95\columnwidth]{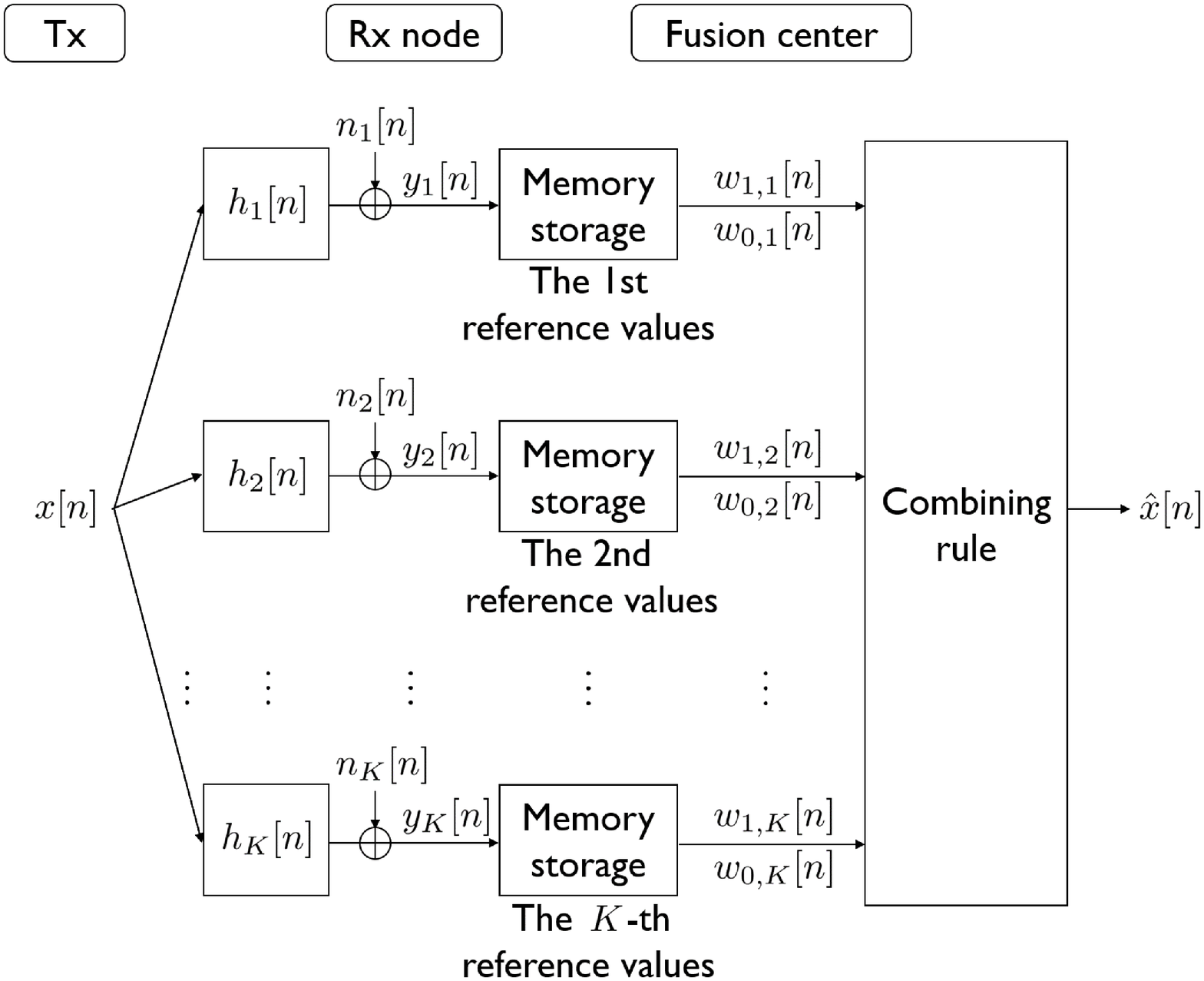}
		\label{data}}
	\caption{The process of OOK symbol detection that is composed of \protect\subref{training} pilot transmissions with $N_p$ time slots and \protect\subref{data} data transmissions.}
	\label{figure2}
\end{figure}\
Fig. \ref{figure2} describes the overall process of pilot and data transmissions.
The pilot symbols employed in this section are the same as in \eqref{training_signal}. Depending on a particular detection criterion, certain reference values, which will be defined as typical or simple statistics like sample averages, are computed at each receiving node using the received pilot signal sequence. During the data transmissions $ N_p+1 \leq n \leq N_p+N_d $, the transmitter sends OOK symbols, and the fusion center computes weights using the instantaneous received data signals and the reference values defined during the pilot transmissions. The fusion center performs OOK symbol detection as
\begin{align}\label{detection}
\hat{x}[n] = \begin{cases}
1, &\text{for } \sum_{k=1}^{K}{w_{1,k}}[n] - \sum_{k=1}^{K}{w_{0,k}}[n]>0, \\
0, &\text{for } \sum_{k=1}^{K}{w_{1,k}}[n] - \sum_{k=1}^{K}{w_{0,k}}[n]<0,
\end{cases}
\end{align}
where $ w_{1,k}[n] $ and $ w_{0,k}[n] $ are the weights for the $ k $-th receiving node, which correspond to how much more likely 1 or 0 are transmitted. The weight difference $ \sum_{k=1}^{K}{w_{1,k}}[n] - \sum_{k=1}^{K}{w_{0,k}}[n] $ has a similar form to the typical binary log-likelihood ratio, but in this framework, the full joint probability distribution is not obtained at the transceivers. Instead, the weights whose definition depends on particular WCNDe are employed. We will explain the three kinds of WCNDe in the following subsections, which employ the reference values based on supervised learning to classify the data signals.

\subsection{Probability-WCNDe (p-WCNDe)}\label{sec4-2}
We first define p-WCNDe, which exploits empirical conditional probabilities that are functions of the pilot symbols. The threshold amplitude, one of the reference values for p-WCNDe, at the $ k $-th receiving node is defined as
\begin{align}\label{mean_all}
A_{\mathrm{th},k} = \frac{1}{N_p}\sum_{m=1}^{N_p} \lvert y_k[m]\rvert.
\end{align}
The $ k $-th receiving node compares the amplitude of each received pilot signal to $A_{\mathrm{th},k}$, and a detected pilot symbol is obtained as
\begin{align}\label{detection_threshold}
\hat{x}_k[n]=
\begin{dcases} 1, & \text{for }\lvert y_k[n] \rvert \geq A_{\mathrm{th},k},\\
0, & \text{for }\lvert y_k[n] \rvert<A_{\mathrm{th},k},\end{dcases}
\end{align}
for $ n=1,\dotsc,N_p $. The $ k $-th receiving node counts how many pilot symbols $ \hat{x}_k[n] $ are correctly detected for $ x[n]=1 $ and $ x[n]=0 $, computed as 
\begin{align}\label{csum1}
g_{1,k}=\sum_{m=1}^{N_p/2}\delta_{\hat{x}_k[m],x[m]},
\end{align}
and
\begin{align}\label{csum0} g_{0,k}=\sum_{m={N_p}/{2}+1}^{N_p}\delta_{\hat{x}_k[m],x[m]},
\end{align}
where $ \delta_{\hat{x}_k[m],x[m]} $ is the Kronecker delta function defined as
\begin{align}\label{kro_delta}
\delta_{\hat{x}_k[m],x[m]}=
\begin{dcases}
1, & \text{for }\hat{x}_k[m]=x[m],\\
0, & \text{for }\hat{x}_k[m]\neq x[m].
\end{dcases}
\end{align}

For the pilot symbols $ x[m] $ for $ m=1,\dots,N_p/2 $, if the pilot symbols  are perfectly detected for some $ k $, the function $ g_{1,k} $ becomes $ N_p/2 $. In contrast, $ g_{1,k} = 0 $ indicates that the entire pilot symbols are incorrectly detected as $ \hat{x}_k[n] = 0 $. This extreme case is possible when the transmit power is extremely low. The function $ g_{0,k} $ can have the same extreme values in similar situations. The $ k $-th receiving node computes two empirical conditional probabilities, another reference values, written as
\begin{align}\label{cond_prob1}
P_{(1\mid 1),k}=\begin{dcases}
\frac{2}{N_p}, & \text{for } g_{1,k}=0, \\
1-\frac{2}{N_p},& \text{for } g_{1,k}=\frac{N_p}{2},\\
g_{1,k}\frac{2}{N_p},& \text{elsewhere},
\end{dcases}
\end{align}
and
\begin{align}\label{cond_prob0}
P_{(0\mid 0),k}
=\begin{dcases}
\frac{2}{N_p}, & \text{for } g_{0,k}=0, \\
1-\frac{2}{N_p},& \text{for } g_{0,k}=\frac{N_p}{2},\\
g_{0,k}\frac{2}{N_p},& \text{elsewhere}.
\end{dcases}
\end{align}
The conditional probability $ P_{(1\mid 1),k} $ corresponds to the event of $ \hat{x}_k[n]=1 $ given $ x[n]=1 $, and $ P_{(0\mid 0),k} $ is similarly defined. The reason not to assign the value 0 or 1 to these empirical probabilities will be explained later.

During the data transmissions, the transmitter transmits an equally likely data symbol $ x[n] $. By measuring the amplitude of the received data signal, each receiving node performs symbol detection as in \eqref{detection_threshold} and computes the two weights using the empirical probabilities in \eqref{cond_prob1} and \eqref{cond_prob0} as
\begin{align}\label{weight_prob1}
w_{1,k}^p[n] = \begin{dcases}
\log{P_{(1\mid 1),k}}, &\text{for } \hat{x}_k[n]=1, \\
\log{(1-P_{(1\mid 1),k})}, &\text{for } \hat{x}_k[n]=0,
\end{dcases}
\end{align}
and
\begin{align}\label{weight_prob0}
w_{0,k}^p[n] = \begin{dcases}
\log{(1-P_{(0\mid 0),k})}, &\text{for } \hat{x}_k[n]=1, \\
\log{P_{(0\mid 0),k}}, &\text{for } \hat{x}_k[n]=0.
\end{dcases}
\end{align}
The fusion center collects all the weights $ w_{1,k}^p[n] $ and $ w_{0,k}^p[n] $ from the $ K $ receiving nodes and finally detects the symbol $ \hat{x}^p[n] $ by combining the weights as in \eqref{detection}. The detection rule of p-WCNDe resembles the conventional LRT that is performed in a binary communication channel \cite{Cover:2006}. The difference is that p-WCNDe utilizes empirical probabilities.

\textbf{Remark 1:} With the finite $ N_p $ pilot symbols, the empirical probability $ P_{(1\mid 1),k} $ can approach one when the transmit power is high. By adopting $ P_{(1\mid 1),k}=1 $, however, the fusion center is not able to consider the possibility that the detection performed at the $ k $-th receiving node is incorrect. The situation of incorrect detection is always possible even at high transmit power because the channel gain changing every time slot can become very small. There is a similar problem when $ P_{(1\mid 1),k}$ goes to zero with low transmit power. Defining bounded values of the empirical probabilities intuitively implies how much trust we have in the detection for each channel condition. In this paper, we set the bound on the conditional probabilities as $ 2/N_p $ and $ 1-2/N_p $ in \eqref{cond_prob1} and \eqref{cond_prob0}, with the implication that there must be at least one correctly or incorrectly detected symbol at each receiving node.

\textbf{Remark 2:} With high transmit power, the detection and error probabilities in the weights defined in \eqref{weight_prob1} and \eqref{weight_prob0} become the bounded values for all the receiving nodes. In this case, the symbol detection conducted at the fusion center boils down to the majority rule since all the weights are the same regardless of $ k $. 

\subsection{Deviation-WCNDe (d-WCNDe)}\label{sec4-3}
In d-WCNDe, each receiving node compares the differences between the reference values that are computed during the pilot transmissions and the amplitude of an instantaneous data signal at each receiving node. Using the received pilot signal sequence for $ x[n]=1 $ and $ x[n]=0 $, two sample averages are computed as
\begin{align}\label{mean1}
A_{1,k} = \frac{2}{N_p}\sum_{m=1}^{N_p/2} \lvert y_k[m]\rvert,
\end{align}
and
\begin{align}\label{mean0}
A_{0,k} = \frac{2}{N_p}\sum_{m=N_p/2+1}^{N_p} \lvert y_k[m]\rvert,
\end{align}
for each $ k $. These sample averages are exploited as the reference values in d-WCNDe.

Using an instantaneous received data signal and the sample averages $ A_{1,k} $ and $ A_{0,k} $, the weights for each $ k $ in d-WCNDe are computed as 
\begin{align}\label{weight_dev1}
w_{1,k}^d[n] = \lvert y_k[n] \rvert - A_{1,k},
\end{align}
and
\begin{align}\label{weight_dev0}
w_{0,k}^d[n] = A_{0,k} - \lvert y_k[n] \rvert,
\end{align}
which consider the two asymmetric magnitude distributions for the cases of $ x[n]=1 $ and $ x[n]=0 $ that have the different mean values. For large $  \lvert y_k[n] \rvert $, the weight $ w_{1,k}^d[n] $ has a larger value than $ w_{0,k}^d[n] $, giving more credit on the hypothesis of $ x[n]=1$, and vice versa. The fusion center finally detects the OOK symbol $ \hat{x}^d[n] $ using $ w_{1,k}^d[n] $ and $ w_{0,k}^d[n] $ as in \eqref{detection}.

\textbf{Remark 3:} For the case of a single receiving node, p-WCNDe and d-WCNDe have the same detection result. The detection rule of p-WCNDe in \eqref{detection_threshold} is simplified to identifying whether $ \lvert y_k[n] \rvert $ is larger than $ A_{\mathrm{th},k} $ or not, which becomes the same as the weight calculations of d-WCNDe in \eqref{weight_dev1} and \eqref{weight_dev0}. The two WCNDe techniques, however, may give different detection results with multiple receiving nodes. In p-WCNDe, once the receiving nodes detect the instantaneous data symbols with \eqref{detection_threshold}, symbol detection at the fusion center is conducted with only the fixed empirical probabilities in \eqref{detection} regardless of $ \lvert y_k[n] \rvert $. On the contrary, for d-WCNDe, the fusion center detects the symbol by exploiting the instantaneous value of $ \lvert y_k[n] \rvert $ as in \eqref{weight_dev1} and \eqref{weight_dev0}.

\subsection{Combination-WCNDe (c-WCNDe)}\label{sec4-4}
For the design of c-WCNDe, the various types of processed values with the same received pilot signals are utilized to improve the symbol detection performance. Specifically, all the reference values and the weights developed for p-WCNDe and d-WCNDe are exploited in c-WCNDe. As will be shown in Section \ref{simul}, c-WCNDe is very robust to fast-varying channel conditions in a human body compared to p-WCNDe and d-WCNDe.

During the pilot transmissions, the threshold amplitude $ A_{\mathrm{th},k} $ and the sample averages $ A_{1,k} $ and $ A_{0,k} $ are computed as in \eqref{mean_all}, \eqref{mean1}, and \eqref{mean0} as well as the empirical probabilities $ P_{(1\mid 1),k} $ and $ P_{(0\mid 0),k} $ as in \eqref{cond_prob1} and \eqref{cond_prob0}, respectively. With an instantaneous received signal during the data transmissions, the weights in c-WCNDe for each $ k $ are computed as
\begin{align}\label{weight_com1}
w_{1,k}^c[n] = 
w_{1,k}^{c1}[n] +w_{1,k}^{c2}[n] ,
\end{align}
and
\begin{align}\label{weight_com0}
w_{0,k}^c[n] = 
w_{0,k}^{c1}[n] +w_{0,k}^{c2}[n],
\end{align}
where
\begin{align}\label{weight_com11}
w_{1,k}^{c1}[n] = 
-\frac{\lvert w_{1,k}^d[n]\rvert ^2}{A_{1,k}},
\end{align}
\begin{align}\label{weight_com01}
w_{0,k}^{c1}[n] = 
-\frac{\lvert w_{0,k}^d[n] \rvert ^2}{A_{0,k}},
\end{align}
\begin{align}\label{weight_com12}
w_{1,k}^{c2}[n] = \begin{dcases}
\frac{1}{A_{\mathrm{th},k}}\log{P_{(1\mid 1),k}^{\lvert w_{1,k}^d[n]\rvert^2}}, &\text{for } \hat{x}_k[n]=1, \\
\frac{1}{A_{\mathrm{th},k}}\log{(1-P_{(1\mid 1),k})^{\lvert w_{1,k}^d[n]\rvert^2}}, &\text{for } \hat{x}_k[n]=0,
\end{dcases}
\end{align}
and
\begin{align}\label{weight_com02}
w_{0,k}^{c2}[n] = \begin{dcases}
\frac{1}{A_{\mathrm{th},k}}\log{(1-P_{(0\mid 0),k})^{\lvert w_{0,k}^d[n] \rvert ^2}}, &\text{for } \hat{x}_k[n]=1, \\
\frac{1}{A_{\mathrm{th},k}}\log{P_{(0\mid 0),k}^{\lvert w_{0,k}^d[n] \rvert ^2}}, &\text{for } \hat{x}_k[n]=0,
\end{dcases}
\end{align}
and the corresponding weights are defined in \eqref{weight_dev1} and \eqref{weight_dev0}. The fusion center combines $ w_{1,k}^c[n] $ and $ w_{0,k}^c[n] $ for all $ k $ as in \eqref{detection} to detect $ \hat{x}^c[n] $.

Focusing on $ w_{1,k}^c[n] $, the two terms $ w_{1,k}^{c1}[n] $ and $ w_{1,k}^{c2}[n] $ are balanced by normalizing with $ A_{1,k} $ and $  A_{\mathrm{th},k} $, respectively. The scalar $ A_{1,k} $ in \eqref{weight_com11} even balances the weights among $ K $ receiving nodes. The second term $ w_{1,k}^{c2}[n] $ is modified to instantaneously adjust the fixed weights (i.e., the empirical probability) of p-WCNDe.  Although heuristic, it is shown in Section \ref{simul} that c-WCNDe outperforms p-WCNDe and d-WCNDe and is comparable to the case of perfect CSI.

\subsection{Discussions on practicality}
To discuss the practicality of the proposed detection techniques, we first briefly compare their computational complexities by counting the number of scalar additions and multiplications. As stated in Section \ref{sec2}, the transmitter sends $ N_p $ symbols for pilot transmissions followed by $ N_d $ symbols for data transmissions. For the marginalization-based approach in Section \ref{sec3}, eLRT computes the likelihood functions in \eqref{eLRT1} and \eqref{eLRT0} during data transmissions, whose complexity is proportional to $ KN_pN_d $. On the contrary, the supervised-learning-based approach computes the reference values during the pilot transmissions and then the weights during the data transmissions for each receiving node, which results in the complexity of same order $ K(N_p+N_d) $ for all WCNDe techniques. Consequently, WCNDe has lower computational complexity than eLRT does. In particular, c-WCNDe can be implemented to compute the reference values and weights in the fusion center with equally low complexity, and only the short length of the received pilot sequence is sufficient, which is examined in Section \ref{simul}.

We now discuss the practicality of WCNDe. Today's commonly used in-body transmitters, e.g., those used in capsule endoscopy, must operate without any external power source. Therefore, the transmitter is required to use minimal power, and one of the simplest solutions is to transmit OOK symbols without any form of hybrid automatic repeat request (HARQ). The receiving nodes may have very limited computational capability to make them cost efficient and just forward the received signals to the fusion center. However, the fusion center, which is located outside of the body with external power source, can have high computing power to process a large amount of data cooperatively, handling data storage, synchronization, error correction, etc. Note that it is not necessary to synchronize the receiving nodes, i.e., the fusion center can post-process and perform WCNDe after collecting all the received signals from the receiving nodes. This function split between the fusion center and receiving nodes makes WCNDe highly practical.

\section{Simulation Results}\label{simul}
\begin{table}[!t]
	\renewcommand{\arraystretch}{1.35}
	\captionsetup{labelsep = newline,justification=centering,font={footnotesize,sc}}
	\caption{Considered distributions of squared channel gain}
	\label{table1}
	\centering
	\begin{tabular}{c c c}
		\hline
		$ d_i(\rho) $ & Distribution model ($ \rho \geq 0 $)& CV \\
		\hline \hline
		$d_1(\rho)$ &Burr $ ([4.71*10^{-7},2.43,5.61]) $&0.4861 \\
		$d_2(\rho)$ &Burr $ ([9.32*10^{-7},3.88*10^1,5.52*10^{-1}]) $& 0.0638 \\
		$d_3(\rho)$ &Burr $ ([2.29*10^{-8},1.21*10^1,5.07*10^{-1}]) $&0.2390 \\
		$d_4(\rho)$ &Burr $ ([5.63*10^{-6},2.40*10^1,3.97*10^{-1}]) $&0.1363\\
		$d_5(\rho)$ &Weibull $ ([1.76*10^{-6},3.88]) $&0.2884 \\
		$d_6(\rho)$ &Burr $ ([3.83*10^{-7},7.06,1.26]) $&0.2392 \\
		$d_7(\rho)$ &Burr $ ([1.31*10^{-6},5.25,1.47]) $&0.3055 \\
		$d_8(\rho)$ &Weibull $ ([1.01*10^{-6},4.05]) $ &0.2774 \\
		$d_9(\rho)$ &Burr $ ([7.76*10^{-6},9.71,7.87]) $&0.1295 \\
		\hline
	\end{tabular}
\end{table}
\begin{figure}
	\centering
	\subfloat[][]{\includegraphics[width=1\columnwidth]{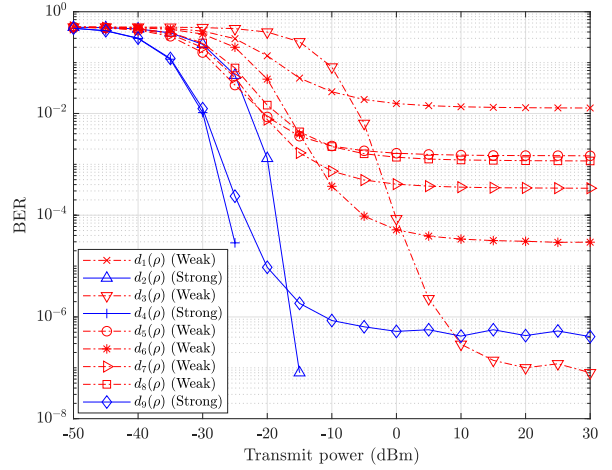}
		\label{figure3a}}
	\hfil
	\subfloat[][]{\includegraphics[width=1\columnwidth]{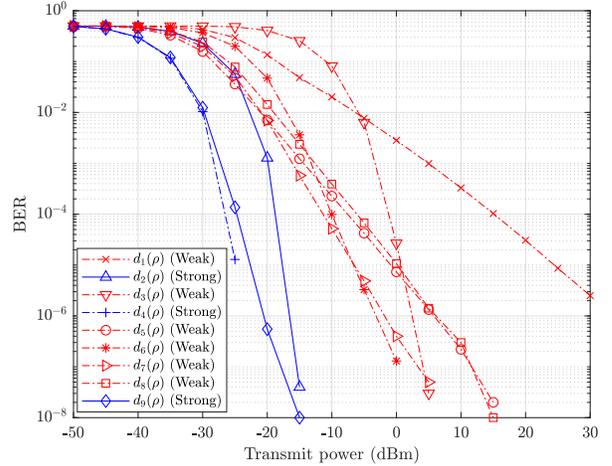}
		\label{figure3b}}
	\caption{BERs for p-WCNDe in \protect\subref{figure3a} and c-WCNDe in \protect\subref{figure3b} following nine different channel distributions between the transmitter and a single receiving node.}
	\label{figure3}
\end{figure}
In this section, we evaluate performance using the uncoded bit error rate (BER) for the two detection approaches developed in Sections \ref{sec3} and \ref{sec4}. For simulations, we adopt the probabilistic models of squared channel gain, i.e., $ \lvert h_k[n]\rvert^2 $, of a human body as in Table \ref{table1}, which are from \cite{Razavi:2019}. We assume the phase of each channel is uniformly distributed from $ 0 $ to $ 2\pi $. In each model, we state the coefficient of variation (CV), which shows how much the channel realization fluctuates in time, defined as $ \sigma_k/\mu_k $ where $ \mu_k $ and $ \sigma_k^2 $ are the mean and variance of the corresponding squared-channel-gain distribution. We set the noise spectral density and the bandwidth to $ N_0=-174$ dBm/Hz and $ B=100 $ MHz, respectively. We assume that the channel model between the transmitter and each receiving node does not vary during the time of interest. The transmitter and each receiving node do not know the corresponding channel distribution and its instantaneous channel value that changes in every time slot, making the system noncoherent. For all simulation results except the last one in Fig. \ref{figure8}, a fixed-length pilot sequence is employed as $ N_p = 40 $, which makes the training overhead in time quite negligible for the $ B = 100 $ MHz bandwidth system. The constant $ c $ for eLRT is set to $ 1/(N_0B) $ in Watt scale, which is a large value, to verify \textbf{Lemma 1}.

For p-WCNDe and c-WCNDe, Fig. \ref{figure3} shows the performance where nine single-receiving-node situations for $ K=1 $ are considered, i.e., $ d_i(\rho) $ for each $ i=1,\cdots,9 $ is chosen as the squared gain distribution. The two WCNDe techniques utilizing a single receiving node make all different performance depending on the channel condition. The BERs of p-WCNDe show an error floor with high transmit power, depicted in Fig. \ref{figure3a}. Reference values in p-WCNDe are less affected by the noise as transmit power increases. However, the reference values, which remain fixed to the bounded value (e.g., $ 1-{2}/{N_p} $), do not cope with the instantaneous channel realization in each time slot as explained in \textbf{Remark 1}. On the contrary, the BERs of c-WCNDe in Fig. \ref{figure3b} decrease without any bound, which verifies robust transmissions with high transmit power.

Meanwhile, we classify the nine channels into two groups that depend on whether the CV of each channel is relatively large or small. Seeing Fig. \ref{figure3}, the dash-dot line corresponds to a weak channel group that has large CVs, and the solid line corresponds to a strong channel group with small CVs. Strong channels tend to have the BERs that start decreasing from small transmit power or achieve lower BERs with high transmit power. The strong channels follow probability distributions with large mean or small variance, allowing reliable communications through the channels with small randomness, and vice versa.

\begin{figure}
	\centering
	\includegraphics[width=1\columnwidth]{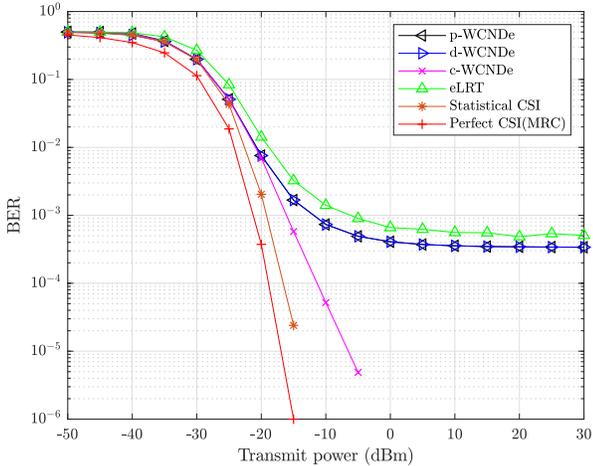}
	\caption{BERs with $ K = 1 $ in which $ d_7(\rho) $ is exploited for perfect, statistical, and no CSI cases.}\label{figure4}
\end{figure}
Focusing on $ d_7(\rho) $ with $ K = 1 $, we discuss the disadvantage of noncoherent detection in Fig. \ref{figure4} with the performance of all proposed techniques in Sections \ref{sec3} and \ref{sec4}. For comparison, we employ coherent detection with perfect CSI where the fusion center performs maximum ratio combining (MRC). The figure also shows the performance with statistical CSI where the symbol detection is performed as explained in \eqref{marg}. MRC shows the best performance followed by the one using statistical CSI while both schemes are impractical for WBANs. As discussed in \textbf{Remark 3} in Section \ref{sec4-3}, p-WCNDe and d-WCNDe show the same performance. While it does not approach the performance with statistical CSI, eLRT also shows an error floor with high transmit power due to the short length of pilot sequence $ N_p $. Resolving the error-floor problem, c-WCNDe outperforms the other noncoherent detection techniques with high transmit power. 

\begin{figure}
	\centering
	\subfloat[][]{\includegraphics[width=1\columnwidth]{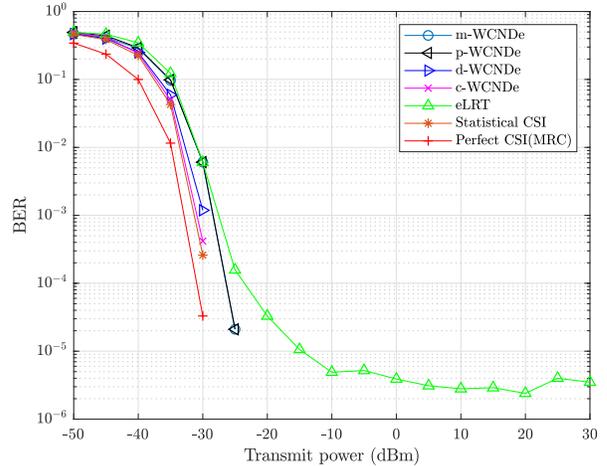}
		\label{figure5a}}
	\hfil
	\subfloat[][]{\includegraphics[width=1\columnwidth]{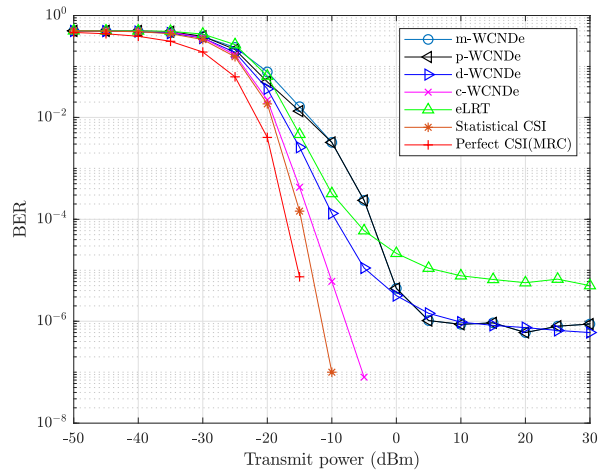}
		\label{figure5b}}
	\caption{BERs for each of two groups that includes three \protect\subref{figure5a} strong and \protect\subref{figure5b} weak channels.}
	\label{figure5}
\end{figure}

In Fig. \ref{figure5}, using one of the channel groups classified as in Fig. \ref{figure3}, we plot the BERs for two extreme cases. Fig. \ref{figure5a} shows the performance exploiting the three strong channels of $ d_i(\rho) $ for $ i=2,4,9 $ with small CVs. However, the optimistic case where the entire channels have small randomness might seldom happen in practice. In Fig. \ref{figure5b}, we plot the performance when using three receiving nodes whose corresponding channels follow the distributions of $ d_i(\rho) $ for $ i=1,3,6 $ in the weak channel group. This channel situation represents a pessimistic case. In addition to the BERs for the three WCNDe variations presented in the previous sections, this figure also compares majority-WCNDe (m-WCNDe) for the noncoherent system. Explaining m-WCNDe briefly, during the pilot transmissions, $ \hat{x}_k[n] $ is computed for each $ k $ using $ A_{\mathrm{th},k} $ as in p-WCNDe. Afterwards, the fusion center detects the data symbol with the majority rule, i.e., by comparing the numbers of the symbol detected as 1 and 0 at each receiving node.

In Fig. \ref{figure5b}, m-WCNDe approaches to p-WCNDe with high transmit power, as discussed in \textbf{Remark 2}. Meanwhile, d-WCNDe shows lower or comparable BER performance to p-WCNDe in both sub-figures of Fig. \ref{figure5}. The weights are computed with continuous values, and the fusion center can conduct more precise detection than that employing the discretized empirical probabilities. Error floors still appear in m-WCNDe, p-WCNDe, d-WCNDe, and eLRT, indicating that these detection techniques do not cope with the noncoherent system. In addition, eLRT underperforms some WCNDe techniques due to the short length of pilot sequence. On the contrary, c-WCNDe outperforms other WCNDe without an~error~floor.

\begin{figure}
	\centering
	\includegraphics[width=1\columnwidth]{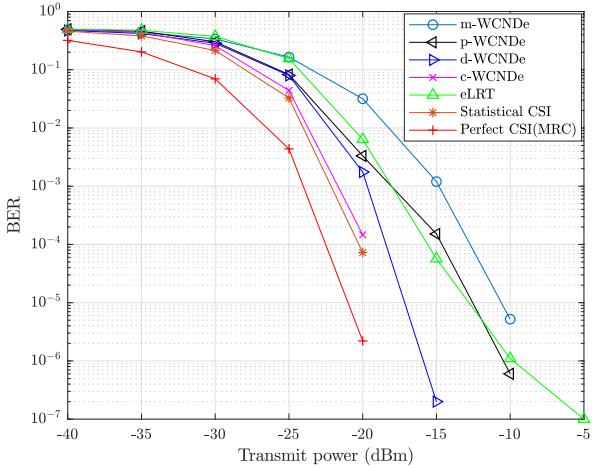}
	\caption{BERs under the various channel distribution models with $ K = 9 $.}\label{figure6}
\end{figure}

In Fig. \ref{figure6}, we plot the performance when employing the entire channels in Table \ref{table1}, which is a definitely more realistic situation with various channel conditions. Comparing to the case using only the strong channels in Fig. \ref{figure5a}, m-WCNDe, p-WCNDe, and d-WCNDe become worse in certain range of transmit power even using more receiving nodes due to the effect of weak channels. Without such degradation, eLRT certainly benefits from utilizing many receiving nodes, and again, c-WCNDe is shown to be the best detection technique among the proposed ones. With only $ N_p=40 $ for the length of pilot sequence, c-WCNDe is comparable to the detection technique with statistical CSI and close to optimal combining using perfect CSI.

\begin{figure}
	\centering
	\subfloat[][]{\includegraphics[width=1\columnwidth]{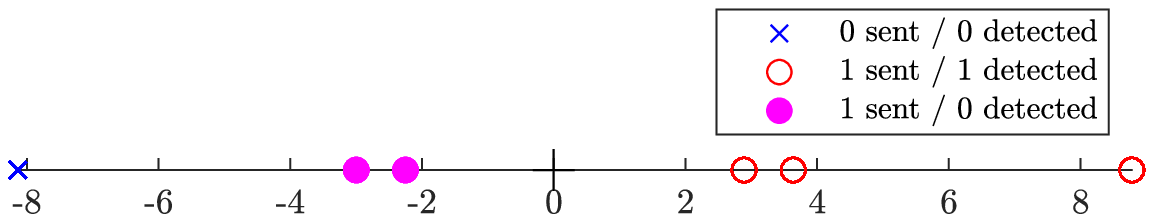}
		\label{figure7a}}
	\hfil
	\subfloat[][]{\includegraphics[width=1\columnwidth]{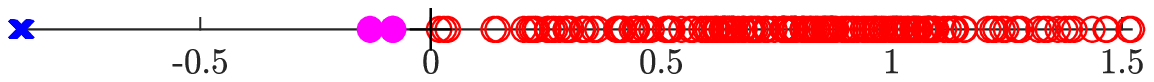}
		\label{figure7b}}
	\hfil
	\subfloat[][]{\includegraphics[width=1\columnwidth]{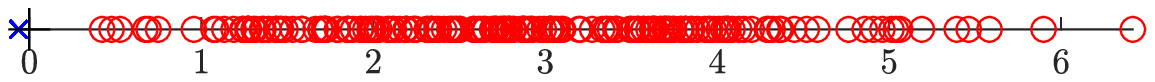}
		\label{figure7c}}
	\caption{Distribution of the normalized weight difference for p-WCNDe in \protect\subref{figure7a}, d-WCNDe in \protect\subref{figure7b}, and c-WCNDe in  \protect\subref{figure7c} with $ K=3 $ channels. Even with high transmit power of $ 40 \text{ dBm} $, p-WCNDe and d-WCNDe suffer from detection error when $ x[n]=1 $ is sent due to the channel fading while c-WCNDe overcomes the fading effectively.}
	\label{figure7}
\end{figure}
Employing multiple receiving nodes, we plot in Fig. \ref{figure7} the distribution of the weight difference, $ \sum_{k=1}^{K}{w_{1,k}}[n] - \sum_{k=1}^{K}{w_{0,k}}[n] $, for the three WCNDe variations in Section \ref{sec4} using received data signals after proper normalization. With the high transmit power of $ 40 \text{ dBm} $, three channels using $ d_i(\rho) $ for $ i=1,5,8 $ are used in this figure where the circles denote for the data symbols of $ x[n]=1 $ and the crosses denote for $ x[n]=0 $. The filled circles indicate the symbols $ x[n]=1 $ detected incorrectly as $ \hat{x}[n]=0 $. Fig. \ref{figure7a} shows the weights have discretized values with empirical probabilities, whereas in Fig. \ref{figure7b} the weights in d-WCNDe are continuous values. When transmit power is high, the distribution for $ x[n]=1 $ has much larger variance than that for $ x[n]=0 $. As shown in the two sub-figures, even with very high transmit power, some data symbols for $ x[n]=1 $ can be detected as $ \hat{x}[n]=0 $ in p-WCNDe and d-WCNDe due to the fast-varying channel. In Fig. \ref{figure7c}, c-WCNDe correctly detects all OOK symbols with high transmit power, resolving the error-floor problem.

\begin{figure}
	\centering
	\includegraphics[width=1\columnwidth]{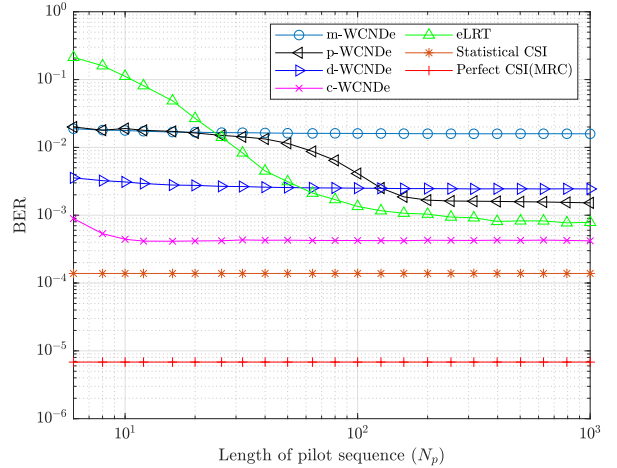}
	\caption{BERs for arbitrary three channels exploiting the length of pilot sequence ranging from 6 to 1000 with fixed transmit power of -15 dBm.}
	\label{figure8}
\end{figure}  
Fig. \ref{figure8} shows the performance with varying length of pilot sequences $ N_p $ from 6 to 1000 with fixed transmit power of $ \-15 \text{ dBm}$. In this figure, we choose three channels of $ d_i(\rho) $ for $ i=1,3,6 $ from Table \ref{table1}. In m-WCNDe and d-WCNDe, there is no extra gain with larger $ N_p $ in which the threshold amplitude and the sample averages as the reference values already converge even with small $ N_p $. On the contrary, the BER of p-WCNDe decreases by increasing $ N_p $ because the empirical probabilities $ P_{(1\mid 1),k} $ and $ P_{(0\mid 0),k} $ become more accurate. The BER of eLRT seems to converge to that of detection using the statistical CSI as derived in \textbf{Lemma 1}, but a large $ N_p $ is required for convergence. The BER of c-WCNDe decreases as $ N_p $ increases, and with only small $ N_p $, c-WCNDe gives better performance than eLRT and the other WCNDe employing the reference values for the noncoherent detection. The results clearly show the practicality of c-WCNDe since its performance approaches that of detection using perfect statistical CSI, even when using a small $ N_p $ value.

\section{Conclusion}\label{conc}
In this paper, we considered a realistic noncoherent system model for WBANs in which a transmitter sends OOK symbols in fast-varying channel conditions. Distributed reception with multiple receiving nodes enables reliable communications where pilot symbols are not used for typical channel estimation. We first developed a marginalization-based approach employing the entire received pilot signals to compute likelihood functions for symbol detection. To mitigate the high complexity of the marginalization-based approach, we also proposed three techniques based on supervised learning that utilize reference values extracted from the received pilot signals and compute weights during data transmissions. The proposed WCNDe techniques enable robust transmissions to various channel conditions for noncoherent WBAN systems. Especially, c-WCNDe achieves outstanding performance due to the well-defined reference values and weights, which also benefits from small pilot overhead. A related future work could be to consider more practical WBAN scenarios such as frequency selective channels or when the transmitter is equipped with multiple transmit antennas.

\appendices
\section{Proof of Lemma 1}\label{proof_conv}
For arbitrary $ k $, we first show the convergence of an argument of the product for the case $ x[n]=1 $, i.e., as $ N_p\rightarrow\infty $ and $ c\rightarrow\infty $,
\begin{align*}
\frac{2}{N_p}\sum_{m=1}^{N_p/2}{\frac{c}{\pi}}e^{-c\lvert y_k[n]-y_k[m]\rvert^2}
\overset{p}{\longrightarrow}
f_{y_k[n]\mid x[n]=1}\bigl(y_k\big\vert x\bigr).
\end{align*}
First, taking $ N_p\rightarrow\infty $, 
\begin{multline*}\label{mm}
\frac{2}{N_p}\sum_{m=1}^{N_p/2}{\frac{c}{\pi}}e^{-c\lvert y_k[n]-y_k[m]\lvert^2} \\
\overset{p}{\longrightarrow}
\bE_{h_k[m],w_k[m]}\left\{{\frac{c}{\pi}}e^{-c\lvert y_k[n]-\sqrt{P} h_k[m] - w_k[m] \lvert^2}\right\},
\end{multline*}
due to the weak law of large numbers and the independence of $ h_k[m] $ and $ w_k[m] $. The expected value above is computed~as 
\begin{align*}
\begin{split}
&\bE_{h_k[m],w_k[m]}\left\{{\frac{c}{\pi}}e^{-c\lvert y_k[n]-\sqrt{P} h_k[m] - w_k[m] \rvert^2} \right\} \\
&=\int_{\mathbb{C}} \int_{\mathbb{C}} {\frac{c}{\pi}}e^{-c\lvert y_k[n]-\sqrt{P} h_k - w \rvert^2} f_{h_k[m]}(h_k) f_{w_k[m]}(w)\mathrm{d}w\mathrm{d}h_k\\
&\stackrel{(a)}{=}\int_{\mathbb{C}} f_{h_k[m]}(h_k){\frac{c}{\pi}}\int_{\mathbb{C}} e^{-c\lvert y_k[n]-\sqrt{P} h_k - w \rvert^2}\frac{e^{-\frac{\lvert w\rvert^2}{N_0B}}}{\pi N_0B}\mathrm{d}w \mathrm{d}h_k
\end{split}\\
\begin{split}
&=\int_{\mathbb{C}} f_{h_k[m]}(h_k){\frac{c}{\pi}}\frac{1}{\pi N_0B}\int_{\mathbb{C}} e^{-c\lvert y_k[n]-\sqrt{P} h_k\rvert^2}\\
&\qquad\times
 e^{-2c\cdot \Re((y_k[n]-\sqrt{P} h_k)w)}e^{-\left(c+\frac{1}{N_0B}\right)\lvert w\rvert^2}\mathrm{d}w \mathrm{d}h_k
\end{split}\\
\begin{split}
&=\int_{\mathbb{C}} f_{h_k[m]}(h_k){\frac{c}{\pi}}\frac{1}{\pi N_0B}\int_{\mathbb{C}} e^{-\frac{c}{1+cN_0B}\lvert y_k[n]-\sqrt{P}h_k\rvert^2}\\
&\qquad\times e^{-\left(c+\frac{1}{N_0B}\right)\left\lvert w-\frac{c}{c+\frac{1}{N_0B}}\left(y_k[n]-\sqrt{P}h_k\right)\right\rvert^2}\mathrm{d}w \mathrm{d}h_k\\
&\stackrel{(b)}{=}\int_{\mathbb{C}} f_{h_k[m]}(h_k){\frac{c}{\pi}}\frac{1}{1+cN_0B}e^{-\frac{c}{1+cN_0B}\lvert y_k[n]-\sqrt{P}h_k\rvert^2} \mathrm{d}h_k,
\end{split}
\end{align*}
where $ (a) $ is based on the fact that $ f_{w_k[m]}(w)=\frac{1}{\pi N_0B}e^{-\frac{\lvert w\rvert^2}{N_0B}} $, and $ (b) $ is due to
\begin{align*}
\int_{\mathbb{C}} e^{-\left(c+\frac{1}{N_0B}\right)\left\lvert w-\frac{c}{c+\frac{1}{N_0B}}\left(y_k[n]-\sqrt{P}h_k\right)\right\rvert^2}\mathrm{d}w=\frac{\pi}{c+\frac{1}{N_0B}}.
\end{align*}
Considering $ c\rightarrow\infty $, it is possible to change the order of the limit and expectation \cite{Bogachev:2007}, which implies
\begin{align*}
\begin{split}
&\underset{c\rightarrow\infty}{\lim}\int_{\mathbb{C}} f_{h_k[m]}(h_k)\\
&\qquad \times {\frac{c}{\pi(1+cN_0B)}}e^{-\frac{c}{1+cN_0B}\lvert y_k[n]-\sqrt{P}h_k\rvert^2} \mathrm{d}h_k \end{split}\\
\begin{split}
&=\int_{\mathbb{C}} f_{h_k[m]}(h_k)\\
&\qquad \times \underset{c\rightarrow\infty}{\lim}{\frac{c}{\pi(1+cN_0B)}}e^{-\frac{c}{1+cN_0B}\lvert y_k[n]-\sqrt{P}h_k\rvert^2} \mathrm{d}h_k
\end{split}\\
\begin{split}
&=\int_{\mathbb{C}} f_{h_k[m]}(h_k)\frac{1}{\pi N_0B}e^{-\frac{\lvert y_k[n]-\sqrt{P}h_k\rvert^2}{N_0B}}\mathrm{d}h_k\\
&=\int_{\mathbb{C}} f_{h_k[m]}(h_k)f_{y_k[n]\mid x[n]=1,h_k[m]}\bigl(y_k\big\vert x,h_k\bigr) \mathrm{d}h_k\\
&=f_{y_k[n]\mid x[n]=1}\bigl(y_k\big\vert x\bigr),
\end{split}\\
\end{align*}
with assumption of independence between the transmit symbol and channel distribution. In a similar way, for the case $ x[n]=0 $ and arbitrary $ k $,
\begin{align*}
\frac{2}{N_p}\sum_{m=N_p/2+1}^{N_p}{\frac{c}{\pi}}e^{-c(y_k[n]-y_k[m])^2}
\overset{p}{\longrightarrow}
f_{y_k[n]\mid x[n]=0}\bigl(y_k\big\vert x\bigr),
\end{align*}
as $ N_p,c\rightarrow\infty $. $ \hfill\ensuremath{\blacksquare} $

\bibliographystyle{IEEEtran}
\bibliography{IEEEtwc}

\end{document}